\documentclass[epj]{svjour}
\usepackage{graphics}
\usepackage{tikz}
\begin{document}
\title{Spectral weights of doublon in interacting Hofstadter model}
%\subtitle{Do you have a subtitle?\\ If so, write it here}
\author{Tirthaprasad Chattaraj \inst{} % etc
% \thanks is optional - remove next line if not needed
%\thanks{\emph{Present address:} }%
}                     % Do not remove
\offprints{}          % Insert a name or remove this line
\institute{University of British Columbia, Vancouver, B.C. V6T1Z1, Canada }
\date{Received: date / Revised version: date}
% The correct dates will be entered by Springer
%
\abstract{
In this article, two-particle Green's functions are computed for different strengths of interactions for particles in Hofstadter lattices, providing informations on spectral weights of doublons. The calculations are performed for a finite lattice without disorder. The splitting in spectra under the effect of external magnetic field for non-interacting and interacting pairs are compared. 
 % end of PACS codes
} %end of abstract
\maketitle
\section{Introduction}
\label{intro}

In recent years, there has been high interest in implemeting the Hofstadter model \cite{Hofstadter} in optical lattices for neutral atoms \cite{Lin} \cite{Goldman} \cite{Creffield} \cite{Struck}. The Hofstadter model takes into account the effect of external magnetic fields on electrons in lattices by making the hopping amplitude complex. The model is mimicked for neutral atoms by periodic modulation of lattice potentials, which averages to zero force, but produces a complex phase factor on momentum dependent hopping or tunneling amplitudes of atoms in lattices \cite{Struck}. This opens the possibility of simulating \cite{Tai} integer and fractional quantum Hall \cite{Yoshioka}  systems and topological insulators  \cite{Bernevig} in disordered 2D optical lattice systems. The model can be derived by Peierls substitution \cite{Peierls} from the tight binding 2D Hubbard model .

\begin{eqnarray}
\mathcal{H}_{} = \sum_{\langle i j \rangle} e^{-\imath \phi_{ij}} J_{ij} a_i^\dagger a_j + \sum_{i} U_{i}  a_i^\dagger a_j^\dagger a_j a_i.
\end{eqnarray}

\begin{figure}
\centering
\hspace{0.0cm}
\begin{tikzpicture}[scale=1.5]
\draw [fill] (0,0) circle [radius=0.03];
\draw [fill] (1,0) circle [radius=0.03];
\draw [fill] (2,0) circle [radius=0.03];
\draw [fill] (3,0) circle [radius=0.03];
\draw [fill] (4,0) circle [radius=0.03];
\draw [fill] (5,0) circle [radius=0.03];
\draw [fill] (0,1) circle [radius=0.03];
\draw [fill] (1,1) circle [radius=0.03];
\draw [fill] (2,1) circle [radius=0.03];
\draw [fill] (3,1) circle [radius=0.03];
\draw [fill] (4,1) circle [radius=0.03];
\draw [fill] (5,1) circle [radius=0.03];
\draw (0,1) --(5,1);
\draw (1,0) --(1,4);
\draw [fill] (0,2) circle [radius=0.03];
\draw [fill] (1,2) circle [radius=0.03];
\draw [fill] (4,2) circle [radius=0.03];
\draw [fill] (5,2) circle [radius=0.03];
\draw (0,2) --(5,2);
\draw (2,0) --(2,4);
\draw [fill] (0,3) circle [radius=0.03];
\draw [fill] (1,3) circle [radius=0.03];
\draw [fill] (2,3) circle [radius=0.03];
\draw [fill] (3,3) circle [radius=0.03];
\draw [fill] (4,3) circle [radius=0.03];
\draw [fill] (5,3) circle [radius=0.03];
\draw (0,3) --(5,3);
\draw (3,0) --(3,4);
\draw [fill] (0,4) circle [radius=0.03];
\draw [fill] (1,4) circle [radius=0.03];
\draw [fill] (2,4) circle [radius=0.03];
\draw [fill] (3,4) circle [radius=0.03];
\draw [fill] (4,4) circle [radius=0.03];
\draw [fill] (5,4) circle [radius=0.03];
\draw (0,4) --(5,4);
\draw (4,0) --(4,4);
\draw (5,0) --(5,4);
\draw (0,0) -- (5,0) -- (5,4) -- (0,4) -- (0,0);
\draw [blue, <->, ultra thick]  (2.2,2) -- (2.8,2);
\node [blue] at (2.5,2.15) {\large $V$};
\draw [red, ->, ultra thick]  (3,2) -- (4,2);
\draw [green, -> , ultra thick]  (2,2) -- (1,2);
\draw [ ->, ultra thick]  (3,2) -- (3,3);
\draw [ -> , ultra thick]  (3,2) -- (3,1);
\draw [ ->, ultra thick]  (2,2) -- (2,1);
\draw [ -> , ultra thick]  (2,2) -- (2,3);
\draw [blue, fill=blue] (2,2) circle [radius=0.13];
\draw [blue, fill=blue] (3,2) circle [radius=0.13];
\node at (1.85,1.5) {\large $J$};
\node at (3.15,1.5) {\large $J$};
\node at (1.85,2.5) {\large $J$};
\node at (3.15,2.5) {\large $J$};
\node [red] at (3.5,2.15) {\large $J e^{i\phi}$};
\node [green] at (1.5,1.85) {\large $J e^{-i\phi'}$};
\draw [ ->]  (5,0) -- (5.5,0);
\node at (5.65,0) {\large $X$};
\draw [ ->]  (0,4) -- (0,4.5);
\node at (0,4.65) {\large $Y$};
\end{tikzpicture}
\vspace{0.5cm}
\caption{The hopping and interaction terms in 2D Hofstadter model for hardcore bosons. The phases $\phi$, $\phi'$ for hopping terms on $X$ axis depends on lattice site indices of $Y$ axis.}
\label{Hoflat}
\end{figure}
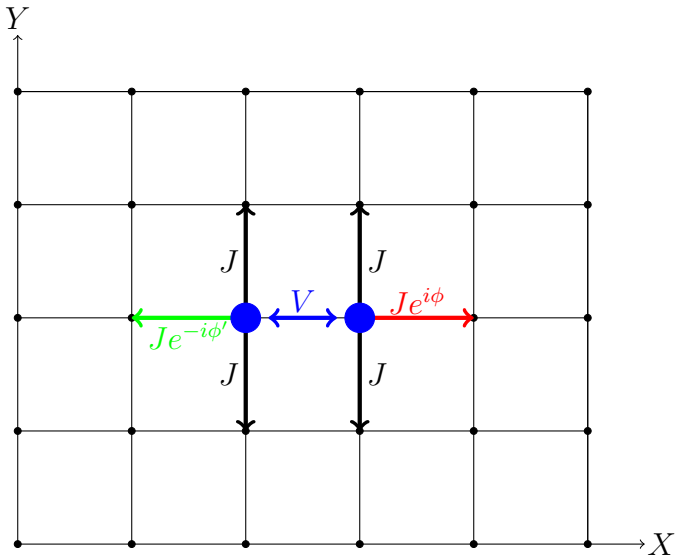

In this paper, the same hamiltonian for two interacting  hard-core bosons is studied. 

\begin{eqnarray}
\mathcal{H}_{pq} &=  \sum_{\langle i j \rangle}\left[ e^{\imath 2\pi \frac{p}{q} i_Y} J a_{i_X +1}^\dagger a_{i_X}  
 + J a_{i_Y}^\dagger a_{i_Y +1} + h.c.  \right] \nonumber\\
& + \sum_{\langle i j \rangle} V  a_i^\dagger a_j^\dagger a_j a_i 
\end{eqnarray}
where $i, j$ are the site indices of two particles and the axes dependency ($X, Y$) is removed from the interaction term for simplicity. The $p$ and $q$ indices are integers coprime to each other. The terms in the Hamiltonian is clearly depicted in Fig. \ref{Hoflat} for a real space 2D lattice. 

The  spectral weight for interacting particles  can be obtained from calculations of two-particle Green's functions as described in Eq. \ref{spec}.  The spectral weight of 2D Hubbard model has been investigated before \cite{Li} \cite{Masanori} \cite{Yang}. However, the effects of external magnetic field on the spectra of interacting particles are yet to be understood \cite{Qin}. The calculations  bring significant difficulty due to the dimensionality and exact diagonalization is limited to small system sizes. A scheme based on recursion \cite{Berciu} \cite{Tirtha} can be employed for such calculations which allows to perform computations for larger system sizes. Using this method of recursion, the spectral weights can be obtained from two-particle Green's functions for a range of interaction. These calculations can provide exact spectral weights for interacting particles from non-interacting regime to strongly interacting regime. However, the two-particle Green's functions do not account for higher order Green's functions which involves three-particle and four-particle interaction terms and so on. Those terms are expected to be negligible where the filling fraction is  not more than two particles per site in the lattices.  

\section{Calculations}
\label{calc}

For the Hofstadter model in 2D lattices, the calculationed are performed using a recursive method from real space retarded Green's functions. 

     \begin{equation}\label{idtt}
              G(\omega) = \frac{1}{E - H + \imath\eta} 
       \end{equation} 
    where $\eta$ is a very small positive real number and  the time-independent propagator from two particles are obtained from $G(m,n, E+\imath\eta) = \langle \mathbf{ m n }| G(E + \imath\eta) | \mathbf{m' n'} \rangle$. Here, $\mathbf{m'}$, $\mathbf{n'}$ are initial sites occupied and $\mathbf{m}$, $\mathbf{n}$ are final sites. The indices $\mathbf{m'}$, $\mathbf{n'}$  are omitted wherever unnecessary for brevity. 

The spectral weight is obtained from 
   \begin{equation}\label{spec}
     A (\mathbf{m' n'}, E) = \frac{-1}{\pi}  \mbox{Im}[G(\mathbf{m' n'}, E + \imath\eta)].
    \end{equation}
These calculations are performed in real space lattices which becomes difficult for large system sizes. Exact diagonalization can only take account of few houndred sites. However, a recursive method developed for such calculations \cite{tirtha} can perform calculations for larger lattices with more than thousand sites. Such calculations provide results with lesser finite size effects. 
The recursions are based on listing several Green's propagators into vectors according to some conserved quantity of the Hamiltonian operation and coupling such vectors with their coupling matrices. 
   \begin{equation}\label{vrec}
 \mathcal{G}_{R =  \mathbf{m}+\mathbf{n}}(E +\imath\eta) = \alpha_R \mathcal{G}_{R-1} + \beta_R \mathcal{G}_{R+1} + \bf C
    \end{equation}
where $\bf C = 0$ ( or $\neq \bf 0$) when $R\neq$ $\mathbf{m'} + \mathbf{n'}$ (or $=\mathbf{ m'} + \mathbf{n'} = R'$).
Once $\alpha_R$ and $\beta_R$ matrices are found for all $R$, the computation becomes straightforward task of recursion as described in detail in reference \cite{Tirtha}. 
In this study, the calculations are performed for interaction strengths between two particles ranging from no interaction ($V = 0$) to weak interaction ($V =  4$) to strong interaction ($V = 12$) cases. The 2D lattice consists of 30 sites per dimension. Several combinations of $(p, q)$ were considered including the case for $q=\infty$ which amounts to zero magnetic field and provides spectral weights of interacting particles for 2D Hubbard model. The combinations (1,2), (1,3), (1,4) are shown in this article. The initial sites $\mathbf{m'}$ and $\mathbf{n'}$ involved in the calculations were taken as two adjacent sites in the middle of the lattice as depicted in Fig. \ref{Hoflat} with complex particles hopping parameters on $X$ axis and real hopping parameters on $Y$ axis.  The value of hopping parameter $J$ is taken as an unit ($J = 1$) for all calculations. The value of $\eta$ was chosen arbitrarily which determines the width of the spectral lines. A standard value of $\eta = 0.05$ was chosen for the calculations of this study which is within the acceptable range ($0.01 \le \eta \le 0.1$) for such studies. The spectral resolution was fixed for these spectra with a uniform gap of $\Delta E = 0.05$. Each separate points of calculation of $E$ within the spectral bandwidth are independent and can be parallelized. The calculations were benchmarked with full diagonalization and the Green's propagators computed using the recursion method are accurate to that computed from full diagonalization \cite{Tirtha} within the single precision range of floating points.

\section{Results}
\label{resu}
%
% For one-column wide figures use
%\begin{figure}
% Use the relevant command for your figure-insertion program
% to insert the figure file.
% For example, with the option graphics use
%\resizebox{0.50\textwidth}{!}{%
 % \includegraphics{2D}
%}
% If not, use
%\vspace{5cm}       % Give the correct figure height in cm
%\caption{Please write your figure caption here}
%\label{fig:5}       % Give a unique label
%\end{figure}
%

% For two-column wide figures use
\begin{figure*}
% Use the relevant command for your figure-insertion program
% to insert the figure file. See example above.
% If not, use
\resizebox{0.998\textwidth}{!}{
  \includegraphics{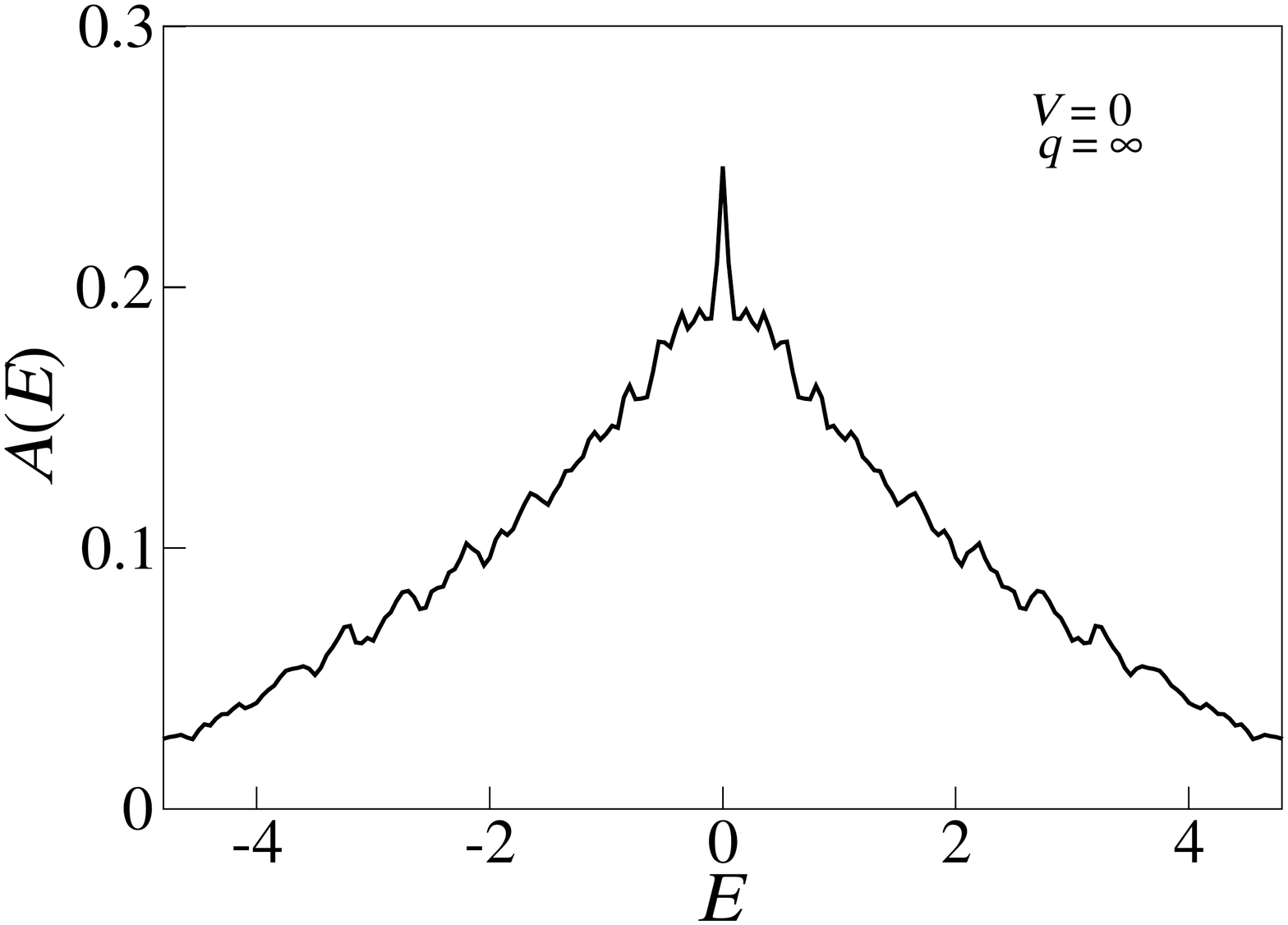}
  \includegraphics{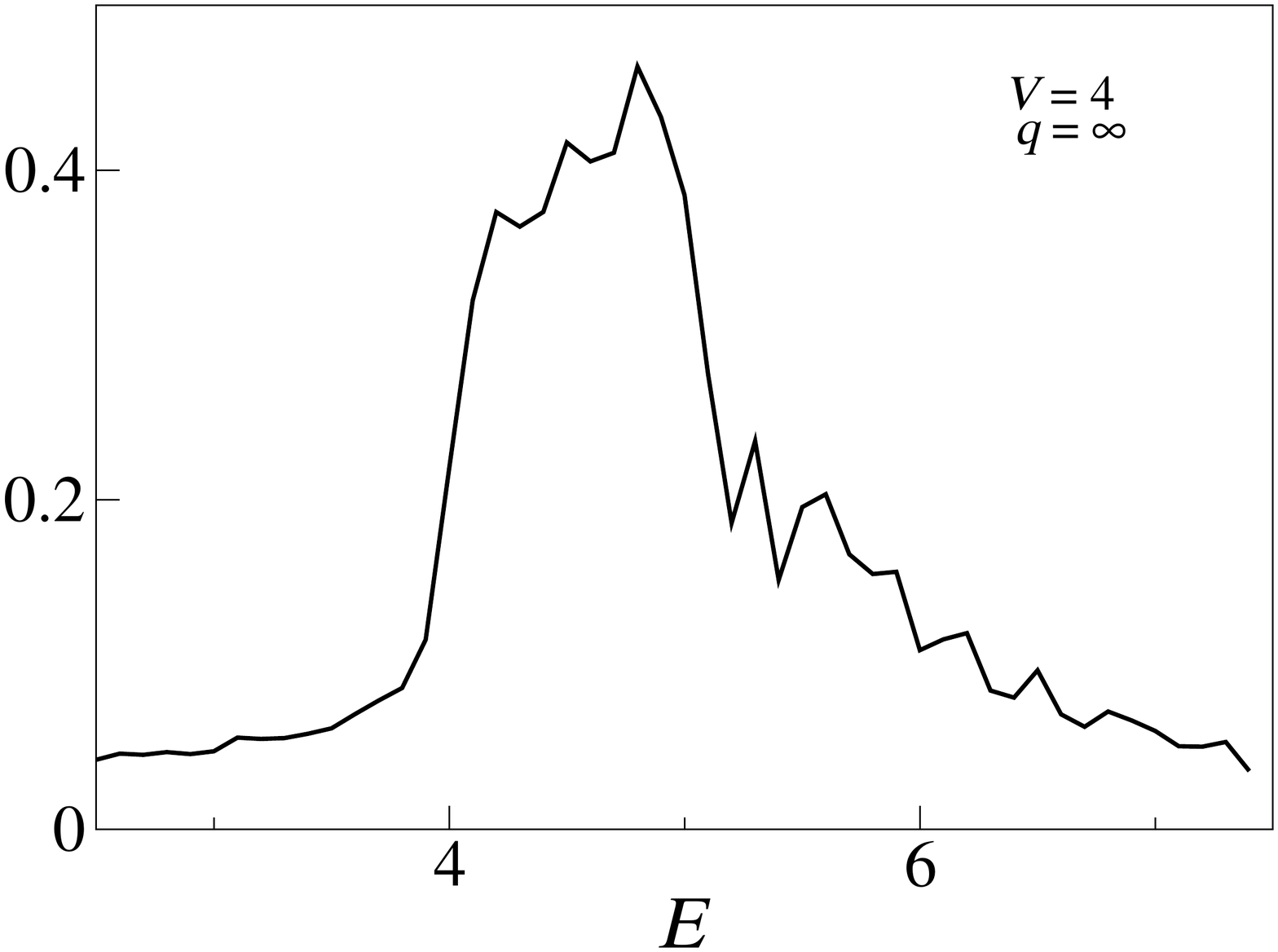}
  \includegraphics{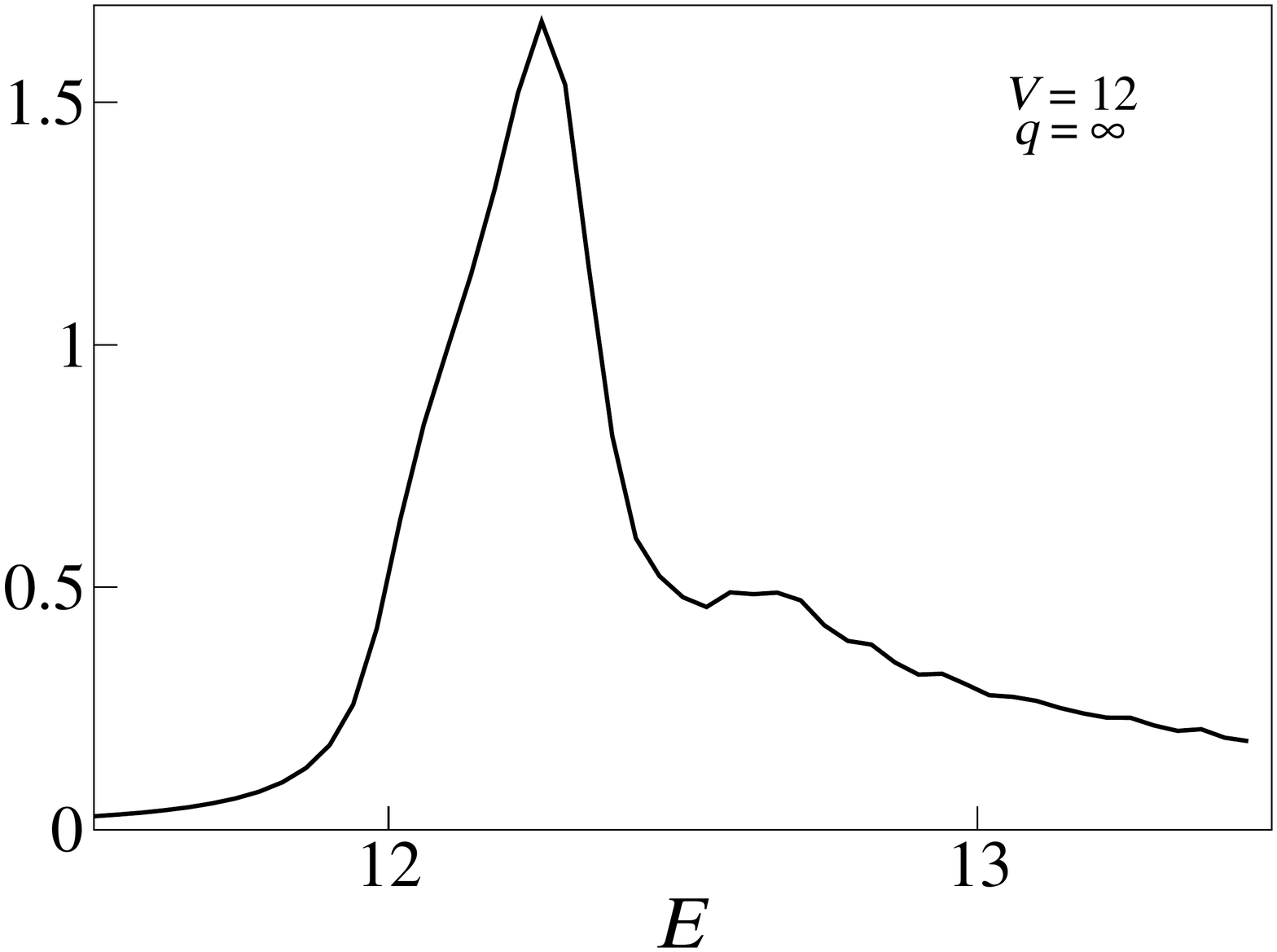}
}
\vspace*{0cm}       % Give the correct figure height in cm
\caption{Spectral weights of doublon for non-interacting ($V = 0$), weakly-interacting ($V = 4$) and strongly-interacting ($V = 12$) cases for $p = 1$ and $q = \infty$}
\label{fig2}       % Give a unique label
\end{figure*}

\begin{figure*}
\resizebox{0.998\textwidth}{!}{
  \includegraphics{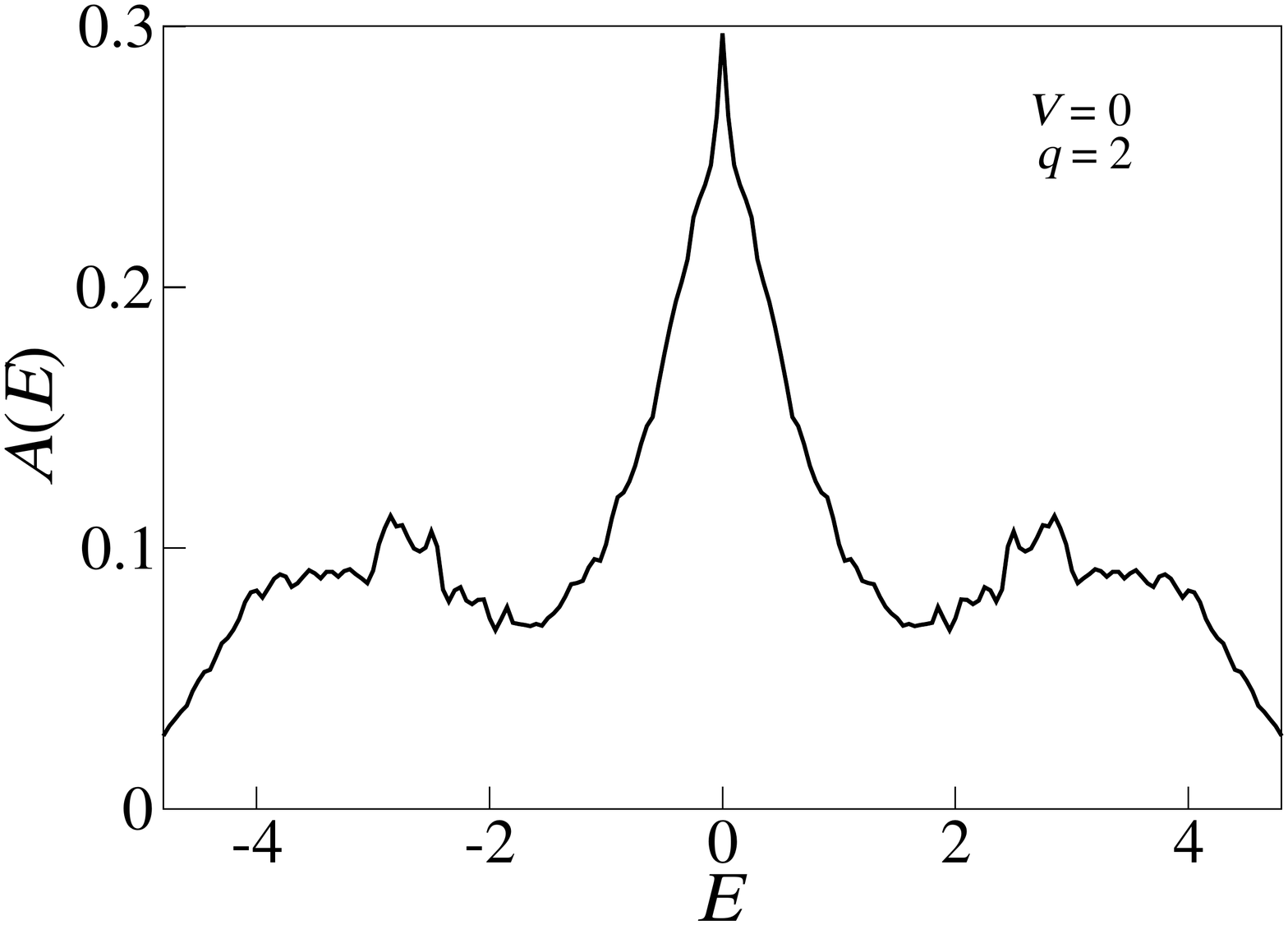}
  \includegraphics{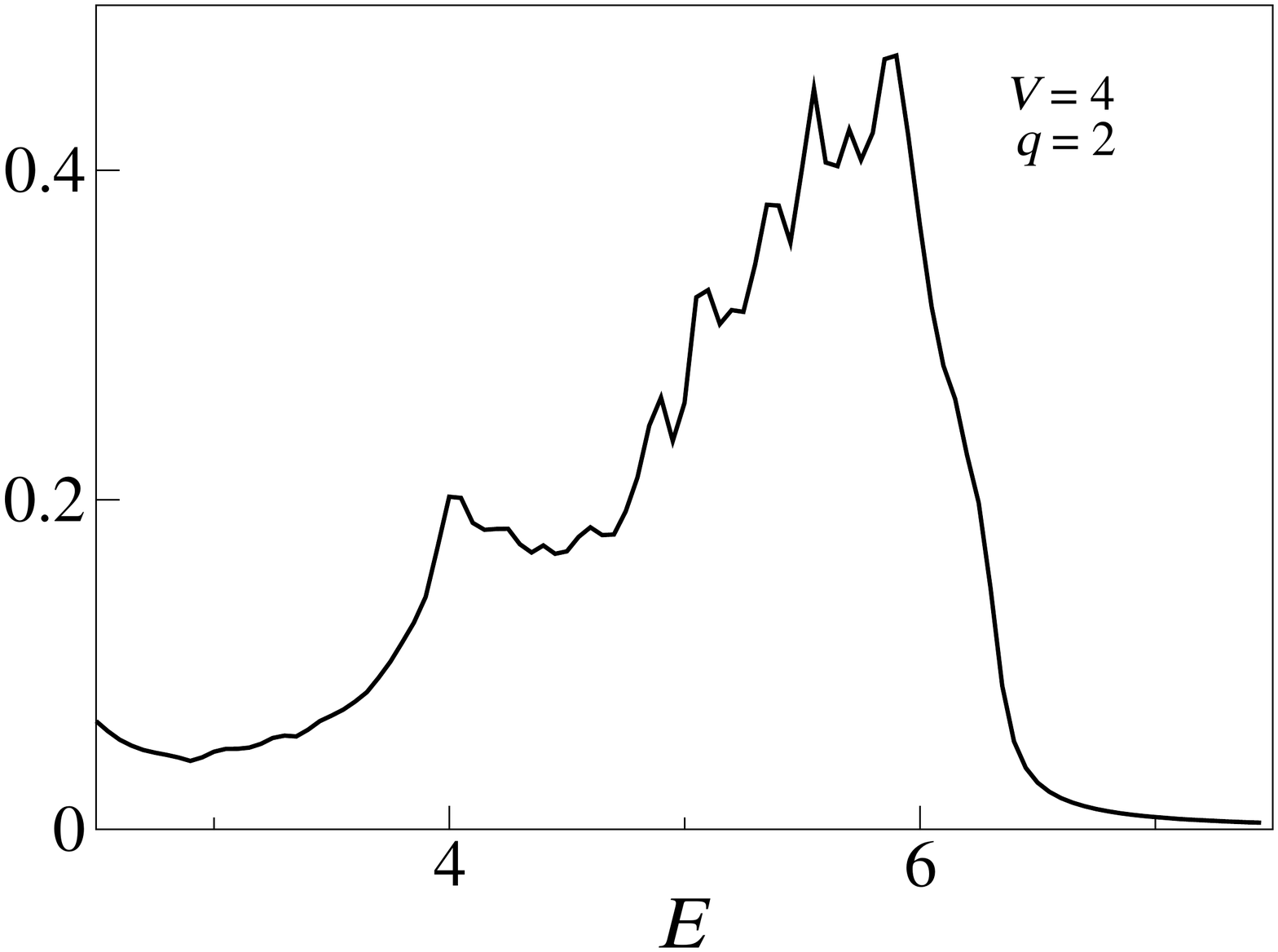}
  \includegraphics{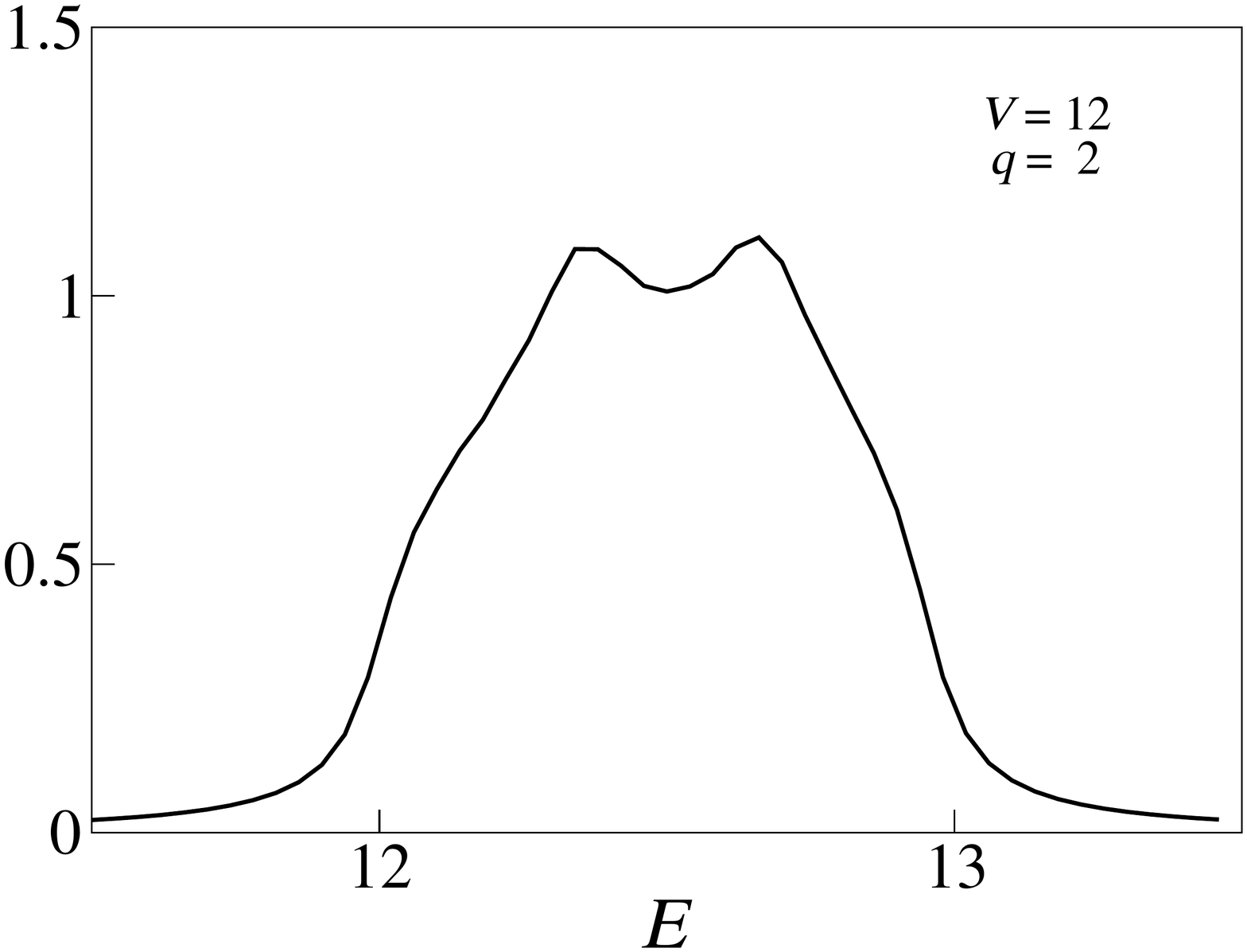}
}
\vspace*{0cm}     
\caption{Spectral weights of doublon for non-interacting ($V = 0$), weakly-interacting ($V = 4$) and strongly-interacting ($V = 12$) cases for $p = 1$ and $q = 2$}
\label{fig3}       % Give a unique label
\end{figure*}

\begin{figure*}
\resizebox{0.998\textwidth}{!}{
  \includegraphics{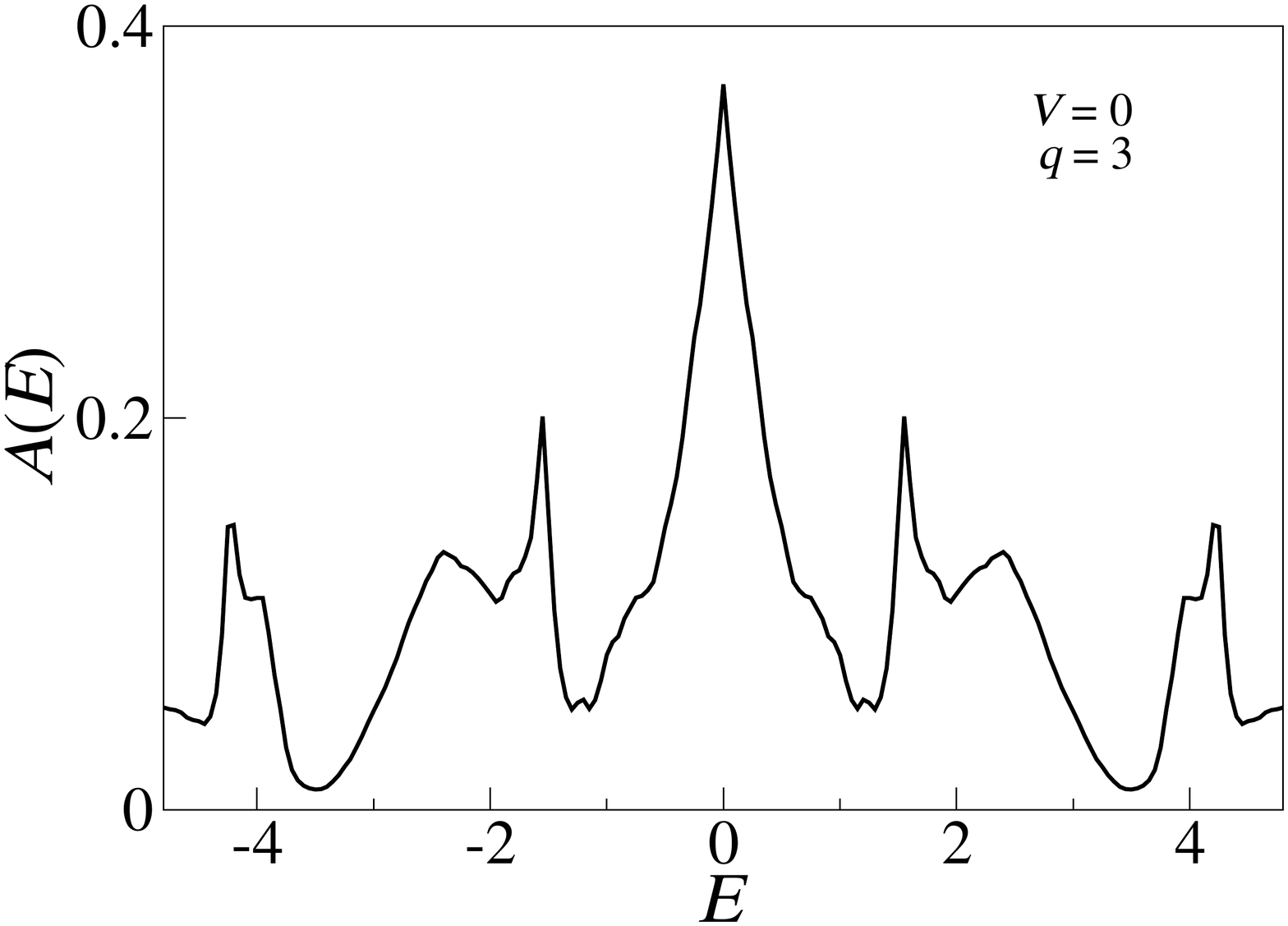}
  \includegraphics{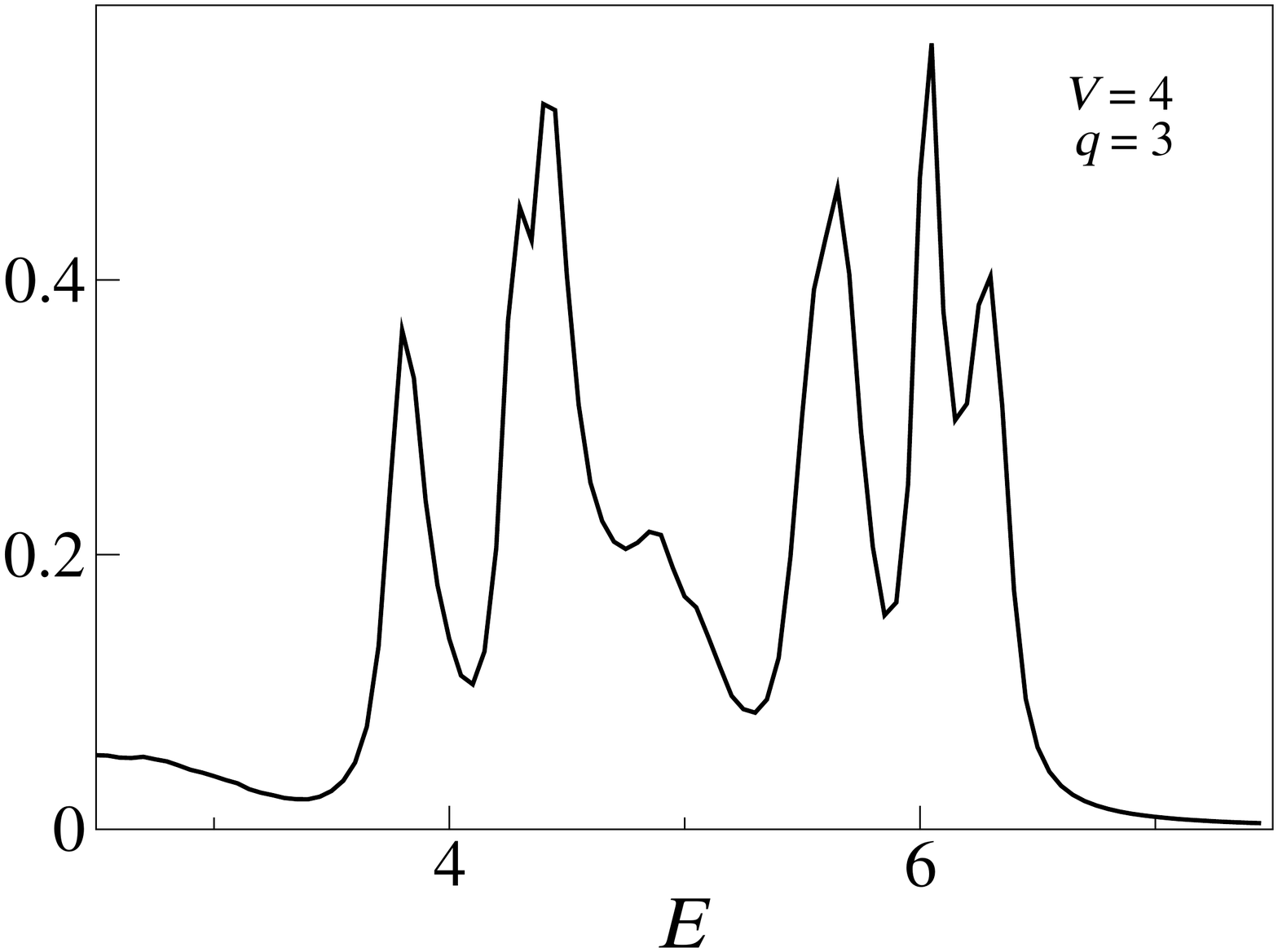}
  \includegraphics{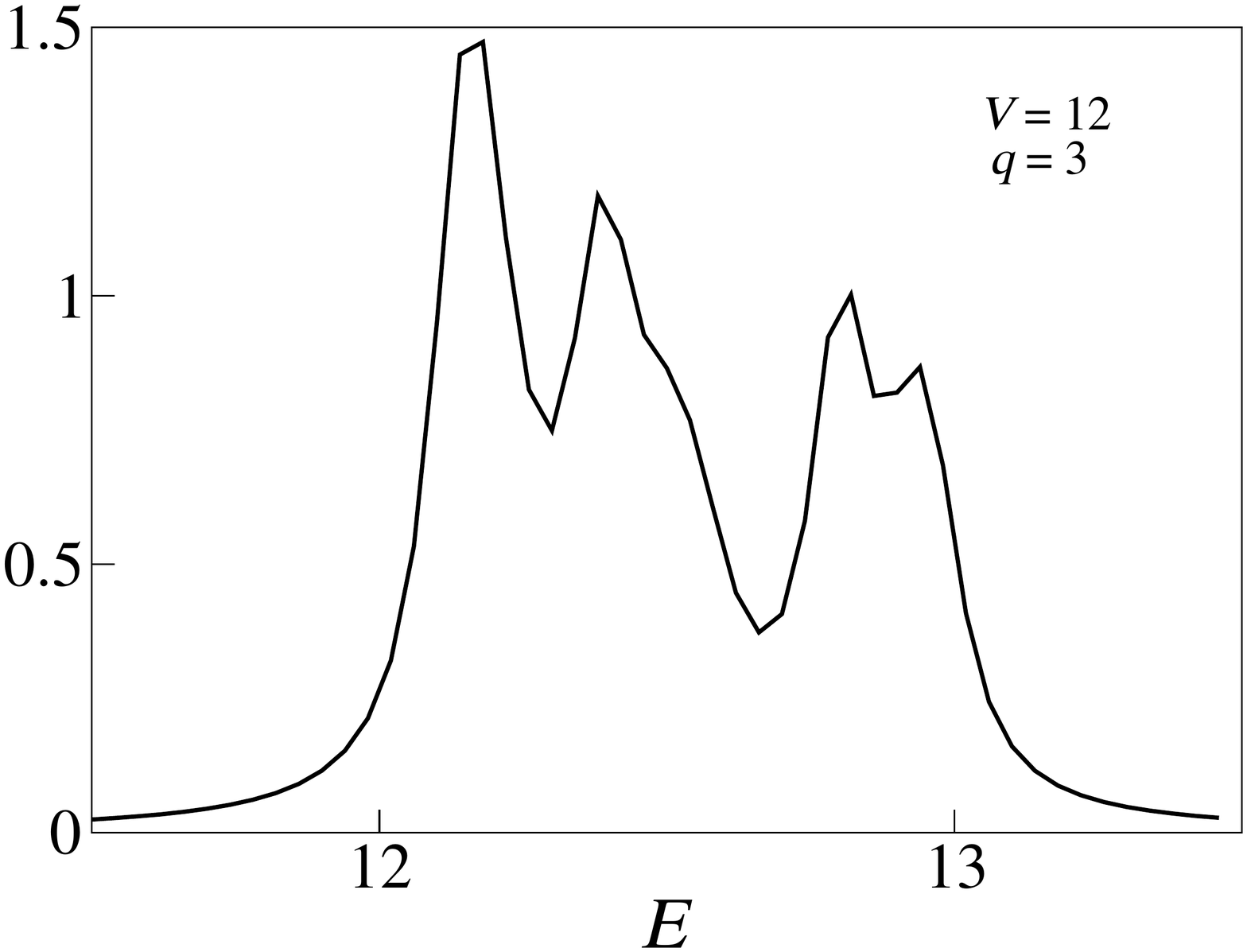}
}
\vspace*{0cm}     
\caption{Spectral weights of doublon for non-interacting ($V = 0$), weakly-interacting ($V = 4$) and strongly-interacting ($V = 12$) cases for $p = 1$ and $q = 3$}
\label{fig4}       % Give a unique label
\end{figure*}

\begin{figure*}
\resizebox{0.998\textwidth}{!}{
  \includegraphics{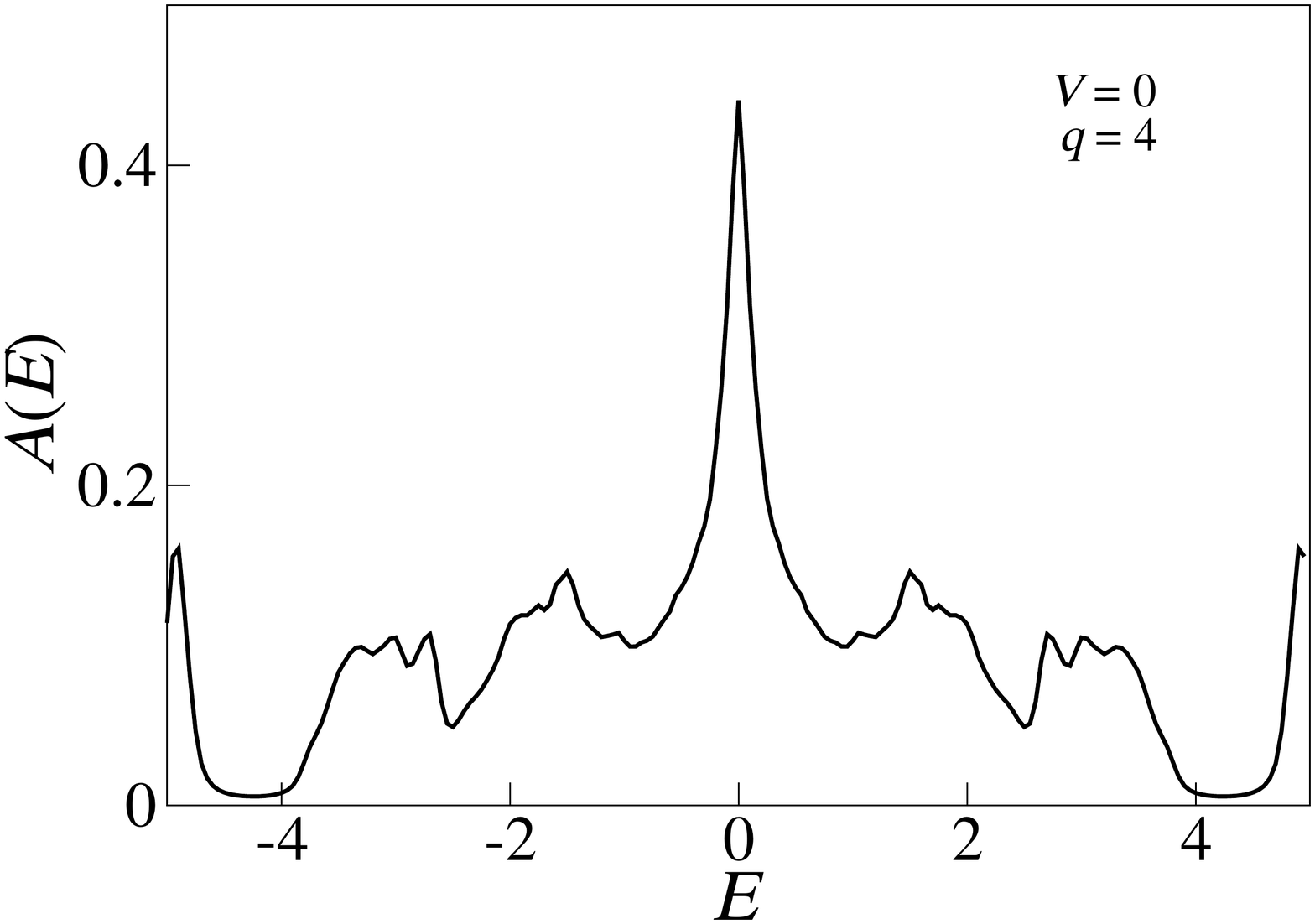}
  \includegraphics{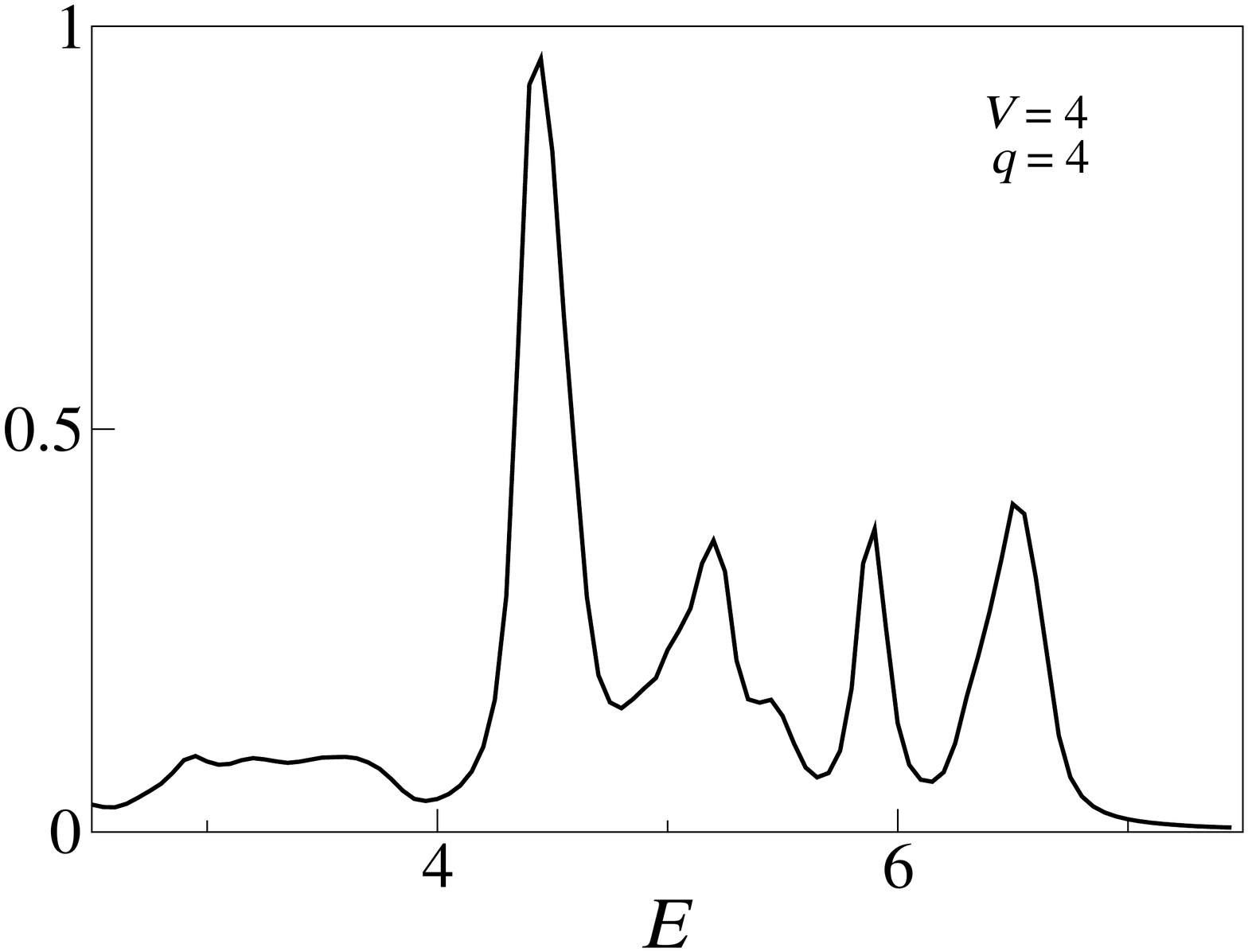}
  \includegraphics{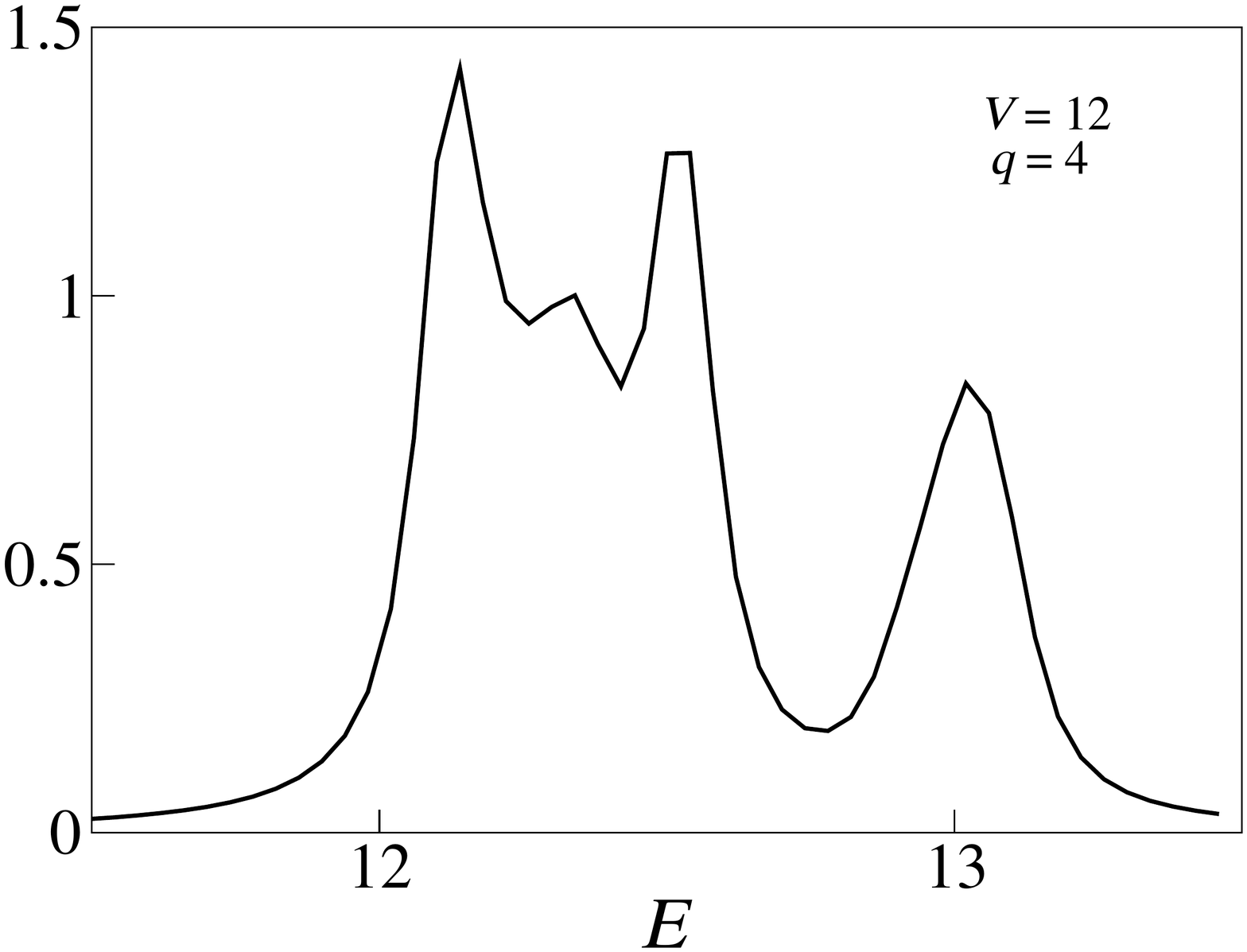}
}
\vspace*{0cm}     
\caption{Spectral weights of doublon for non-interacting ($V = 0$), weakly-interacting ($V = 4$) and strongly-interacting ($V = 12$) cases for $p = 1$ and $q = 4$}
\label{fig5}       % Give a unique label
\end{figure*}

The spectral weights of the doublon without any magnetic field show signatures of binding when interaction strength is increased as shown in Fig. \ref{fig2}. The weak interaction case ($V = 4$) show a sharp increase in weight from $E = 4$ followed by a plateau and a sharp drop at $E = 5$ followed by a long tail. The strong interaction case ($V = 12$) show a sharp increase from $E = 12$ followed by a sharp drop before $E = 12.5$ and a long tail. The reduction in bandwidth of the spectra can be attributed to flattening of bound state band with increase in interaction strength. 

The spectral weights for the cases with $p/q > 0$ show expected splitiing. However the number of major bands which can be observed from the spectra are different for non-interacting case from interacting case. The non-interacting particles show a sharp peak at $E = 0$ with the $q-1$ number of broad peaks on both sides of zero. Each of these broad peaks has a non-homogeneous shape with more peaks inside them.  For increasing interactions, these broads peaks seem to be merging with each other. These spectral weights indicate that the underlying physics of interacting particles in the presence of external magnetic fields is complex and the splitting of spectra has a complex dependence on $q$. The distribution of weight  is symmetrical for non-interacting case. In presence of interaction the the distribution of weight is inhomogeneous across major bands. For ($p=1, q=2$) case, the weak interaction case show larger spectal weight towards the higher energies while for ($p=1, q=4$) the spectra has more weight towards lower energies for the same interaction strength.  Higher $q/p$ show sharper splitting and increased bandwidth. An trend in asymmetric distribution of weight can be noted as higher $q/p$ and higher strength of interaction $V$ effecting in moving the weight to lower energies. This trend can be understood to be reversed for $q/p > 1/2$ as $(q-p)/q$ and $q/p$ has same spectra. However, the effect of splitting on character of bound state cannot be interpreted from the spectral weights.

\section{conclusion}
\label{conc}
In this study, the spectral weight for two interacting particles occupying adjacent sites in an ideal 2D lattice under the effect of homogeneous external magnetic field is obtained from recursive computation of two-particle Green's functions. The effect of external magnetic field on spectra of interacting particles are observed and analyzed. more  calculations, however, are required for understanding the effect on dynamics of doublon.

%
% For tables use
%\begin{table}
%\caption{Please write your table caption here}
%\label{tab:1}       % Give a unique label
% For LaTeX tables use
%\begin{tabular}{lll}
%\hline\noalign{\smallskip}
%first & second & third  \\
%\noalign{\smallskip}\hline\noalign{\smallskip}
%number & number & number \\
%number & number & number \\
%\noalign{\smallskip}\hline
%\end{tabular}
% Or use
%\vspace*{5cm}  % with the correct table height
%\end{table}
%

%
% BibTeX users please use
% \bibliographystyle{}
% \bibliography{}
%
% Non-BibTeX users please use

\end{document}